\documentclass{iau379}

\usepackage{aas_macros}

\usepackage{graphicx}
\usepackage{amsmath}
\usepackage{multirow}

\begin{document}

\lefttitle{Yang et al.}
\righttitle{IAU Symposium 379: Milky Way dwarf galaxies : a recent infall}

\jnlPage{1}{7}
\jnlDoiYr{2023}
\doival{10.1017/xxxxx}

\aopheadtitle{Proceedings of IAU Symposium 379}
\editors{P. Bonifacio,  M.-R. Cioni, F. Hammer, M. Pawlowski, and S. Taibi, eds.}

\title{Gaia EDR3 proper motions, energies, angular momenta of Milky Way dwarf galaxies: a recent infall to the Milky Way halo}

\author{
Yang Y. $^1$,    
Hammer F.$^1$,   
Li H.$^2$,       
Pawlowski M. S.$^3$,  
Wang J. L.$^4$,            
Babusiaux C.$^5$     
Mamon G. A.$^6$,   
Bonifacio P.$^1$,     
Jiao Y.$^1$,          
\& Wang H.$^7$          
}

\affiliation{
$^1$ Observatoire de Paris, Paris Sciences et Lettres, CNRS France\\
$^2$ School of Physical Sciences, University of Chinese Academy of Sciences, Beijing, China\\
$^3$ Leibniz-Institut fuer Astrophysik Potsdam, Germany\\
$^4$ CAS Key Laboratory of Optical Astronomy, National Astronomical Observatories, Beijing 100101, China\\
$^5$ Universit\'{e} de Grenoble-Alpes, CNRS, IPAG, F-38000 Grenoble, France\\
$^6$ Institut d'Astrophysique de Paris, CNRS, France\\
$^7$ Centro Ricerche Enrico Fermi, Roma, Italy
}

\begin{abstract}
Gaia EDR3 has provided proper motions of Milky Way (MW) dwarf galaxies with an unprecedented accuracy, which allows us to investigate their orbital properties. We found that the total energy and angular momentum of MW dwarf galaxies are much larger than that of MW K-giant stars, Sagittarius stream stars and globular clusters. It suggests that many MW dwarf galaxies have had a recent infall into the MW halo. We confirmed that MW dwarf galaxies lie near their pericenters, which suggests that they do not behave like satellite systems derived from Lambda-Cold-Dark-Matter cosmological simulations. These new results require revisiting the origin of MW dwarf galaxies, e.g., if they came recently, they were likely to have experienced gas removal due to the ram pressure induced by MW's hot gas, and to be affected by MW tides. We will discuss the consequences of these processes on their mass estimation.
\end{abstract}


\maketitle

\section{Introduction}

Within a distance of 260 kpc around the Milky Way (MW), there are 46 dwarf spheroidal galaxies (dSphs), including the 11 so-called classical dSphs: Leo I, Leo II, Fornax, Sculptor, Sextans, Carina, Ursa Minor (UMi), Sagittarius (Sgr), Draco, and ultra-faint dwarf galaxies (UFDs), those with extremely low values of central surface brightness. UFDs possess the least stellar mass, which could be only a few hundred solar masses, such as Segue1 \citep{McConnachie2012}. It is worth noting, however, that three UFDs are eventually as massive as the classical dSphs, which are Canes Venatici \citep{Martin2008}, Antlia 2 and Crater 2 \citep{Torrealba2016,Ji2021}
 despite their ultra-faint central surface brightness.

A large effort has been made to search for the signatures of tides from the MW in the classical dSphs, however, apart from Sgr, no firm evidence was found \citep{Walker2013}. Therefore, the dSphs were thought to be long-lived satellites of the MW, with their first infall around 8-10 Gyr ago, as speculated based on their relatively regular morphologies and predominantly old stellar populations. These characteristics are primarily observed in classical dSphs, and largely shared by UFDs \citep{Simon2019}.

The MW dwarf galaxy system shows several peculiar properties that contradict to the prediction of Lambda-Cold-Dark-Matter (LCDM) simulations \citep{Bullock2017}. For example, the missing satellite problem, i.e., the total number of observed dwarf galaxies in the MW is so small, by a factor of at least 200, compared to that predicted by LCDM simulations. To resolve this issue, one has to assume that most of low mass sub-halos are unable to activate their star formation, and thus remain fully dark today.

One of the well-known peculiar properties of the MW dwarf galaxies is the Vast POlar Structure \citep[VPOS,][]{Pawlowski2012} which includes not only the classical dwarf galaxies \citep[previously known as the Disk of Satellites, see][]{Kroupa2005,Lynden-Bell1976}, but also UFDs. This is a thin structure, almost perpendicular to the MW disk, where galaxies show coherent motions and share similar orbital angular momentum directions.

Based on Gaia Data Release 2 (DR2), \citet{Fritz2018} found that not all, but only 60\% of the dwarf galaxies could be in VPOS. With  much  more accurate proper motions from Gaia Early DR3 (EDR3) and an enlarged sample of dSphs, we confirmed their result in \citet{Li2021}.  Surprisingly, we further discovered a new large-scale structure perpendicular to both the VPOS and the MW disk \citep{Hammer2021}. It consists of 20\% (i,e., 10) of dwarf galaxies orbiting or counter-orbiting with the Sgr. This result further complicates our understanding of the MW's dwarf galaxy system, as up to 80\% of the galaxies are preferentially distributed in planes.

VPOS challenges the LCDM model of cosmology because its kinematics-space distribution makes the MW an outlier in LCDM cosmology. Many efforts have been made to explain the special characteristics of the MW's dwarf galaxies within the LCDM frame. Despite various proposed remedies, the MW dwarf galaxy system stands out as atypical when compared to LCDM simulations.

Cosmologists have acknowledged the exceptional situation of the MW dwarf galaxy system when compared to LCDM simulations. Based on Gaia DR2 proper motions, \citet{Cautun2017} found that MW dwarf galaxies show excesses of tangential velocity on their orbits, making the MW exceptional (only 1.5\%) among LCDM satellites systems. Gaia EDR3 proper motions confirmed this conclusion \citep{Hammer2021}.  

Why do MW dwarf galaxies exhibit such peculiar spatial and velocity distributions? Apart from challenging the LCDM cosmology, could we trace their orbital properties and learn about their origins? Ultimately, the dynamics, particularly the orbital history of dwarf galaxies, is the primary reason that drives the assembly of the dwarf galaxies into the MW, as well as their formation and transformation.

\begin{figure}[t]
  \centerline{\vbox to 6pc{\hbox to 10pc{}}}
  \includegraphics[width=13cm]{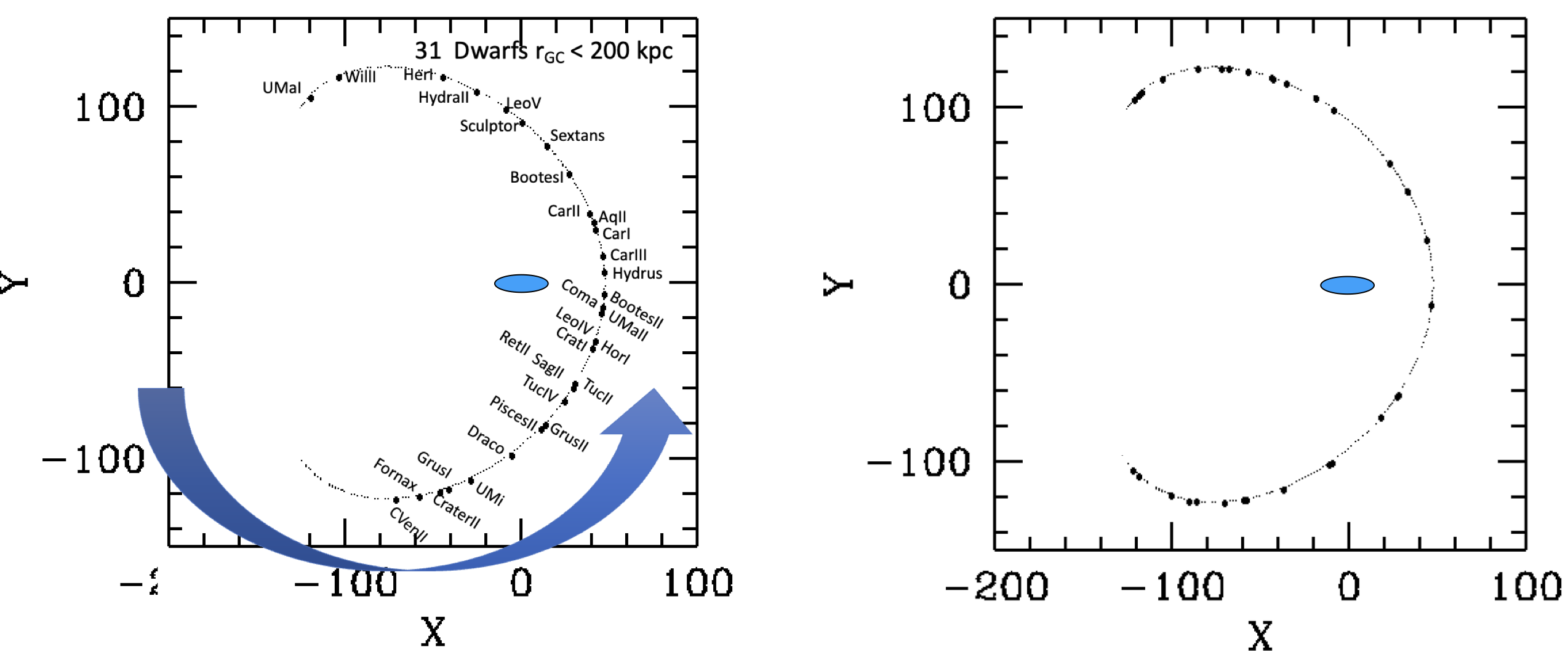}
  \caption{{\it Left}: Orbital phase distribution of 31 dwarf galaxies currently within 200 kpc around MW. The orbital phase of each dwarf galaxy has been normalized to the median orbit of this subsample, i.e., the orbit with an eccentricity of 0.55, assuming a MW total mass of $8.1\times 10^{11}M_{\odot}$ (see text for more details). {\it Right}: Orbital phase distribution assuming a random distribution in time, on the same orbit.}
  \label{fig:rperi}
\end{figure}

\section{Orbital properties of MW dwarf galaxies revealed by Gaia EDR3}
Gaia EDR3 \citep{Gaia2016,Lindegren2021,Riello2021,Gaia2021} provides a significant improvement in the precision of astrometry measurements by a factor $\approx$~2 for individual sources. This helps to reduce the errors in proper motion determination of the MW dwarf galaxies, especially for UFDs because their proper motions are mostly dominated by statistical errors. 
In \citet{Li2021}, we measured the proper motions using Gaia EDR3 for a larger sample of MW dwarf galaxies \cite[46 vs. 39 in][based on Gaia DR2]{Fritz2018}.
After taking into account distance and radial velocity measurements from the literature, we established a complete six-dimensional (6D) dynamical data (3D in space  and 3D in velocity) of each dwarf galaxy in the reference of the MW. This data allows us to study all dynamical aspects of the dSphs' orbits.

\subsection{A strong tendency to be near pericenter}
\label{sec:rperi}
Based on Gaia EDR3 proper motion measurements, it was noticed that MW dwarf galaxies show excesses of tangential velocity on their orbits, which  suggests that MW dwarf galaxies may be preferentially approaching their pericenters or apocenters. To investigate further, we calculated their orbital properties, such as orbital angular momentum, eccentricities, pericenters or apocenters \citep{Li2021}. Given the large uncertainties in determining the MW total mass \citep{Jiao2021}, we performed the calculations by assuming different MW mass profiles, ranging from 2.8 to 15 $\times10^{11}M_{\odot}$, that fit the MW's rotation curve from Gaia DR2 \citep{Eilers2019}. It is not surprising that orbital parameters change with the MW mass, because a higher MW mass gives more circularized solutions, while a lower MW mass gives more eccentric solutions. 

Although we cannot draw conclusions on the exact values of orbital properties, we noticed that MW dwarf galaxies are, overall, on quite eccentric orbits. For instance, for a MW mass of $8.1\times10^{11}M_{\odot}$ \citep{Li2021}, the median eccentricity is 0.55. What is more puzzling is that the galaxies tend to be near to their pericenters, which holds true regardless of the choice of MW masses. With careful analysis, the tendency was found to be significant with a probability of 0.001 for a MW mass of $8.1\times10^{11}M_{\odot}$, when compared to a null hypothesis, i.e., MW dwarf galaxies are randomly distributed in time on their orbits \citep{Li2021}.
This contradicts what is expected for a virialized satellite system, in which satellite galaxies have to be distributed closer to their apocenters, in accordance with Kepler's law (or its generalization for extended mass profile). On an elliptical orbit, an object spends a relatively short amount of time near the pericenter. This is illustrated in Figure.~\ref{fig:rperi}.
Why do MW dwarf galaxies show such a tendency? Could they be coordinated in time to arrive at the MW?

\begin{figure}[t]
  \centerline{\vbox to 6pc{\hbox to 10pc{}}}
  \includegraphics[width=13cm]{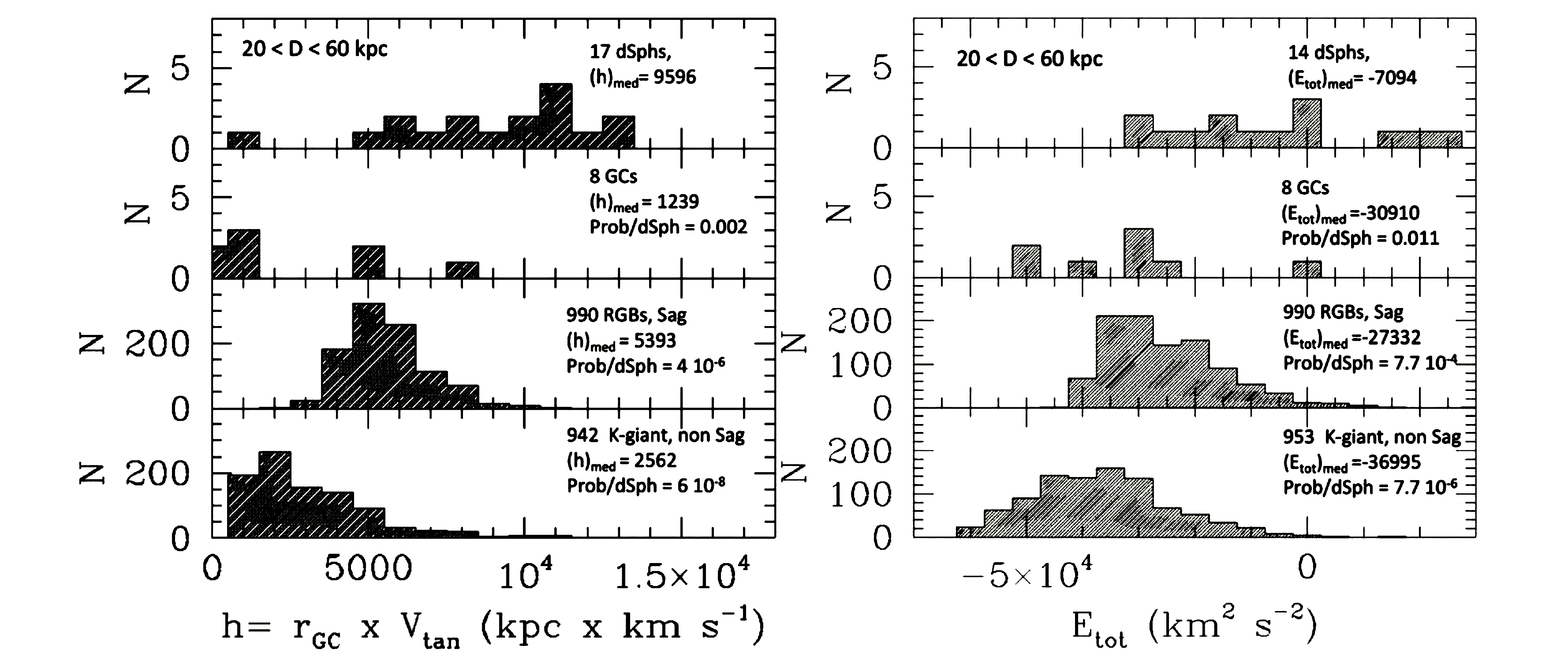}
  \caption{{\it Left}: Comparison of orbital angular momentum ($h$) distribution between (from top to bottom) MW dwarf galaxies,  globular clusters, Sgr Stream stars, and K-giant stars. In each panel, we have indicated the number of objects for each species,  their median $h$ and the probability that the distribution is consistent with that of the dwarf galaxies. All samples are limited to a range of 20 to 60 kpc to ensure completeness.
{\it Right}: The same comparison is made as in {\it Left}:, but for total orbital energy, $E_{\rm tot}$. Two dwarf galaxies are excluded from the $E_{\rm tot}$ comparison due to their large error bar.
  }
  \label{fig:hande} 
\end{figure}

\begin{figure}[t]
  \centerline{\vbox to 6pc{\hbox to 10pc{}}}
  \includegraphics[width=13cm]{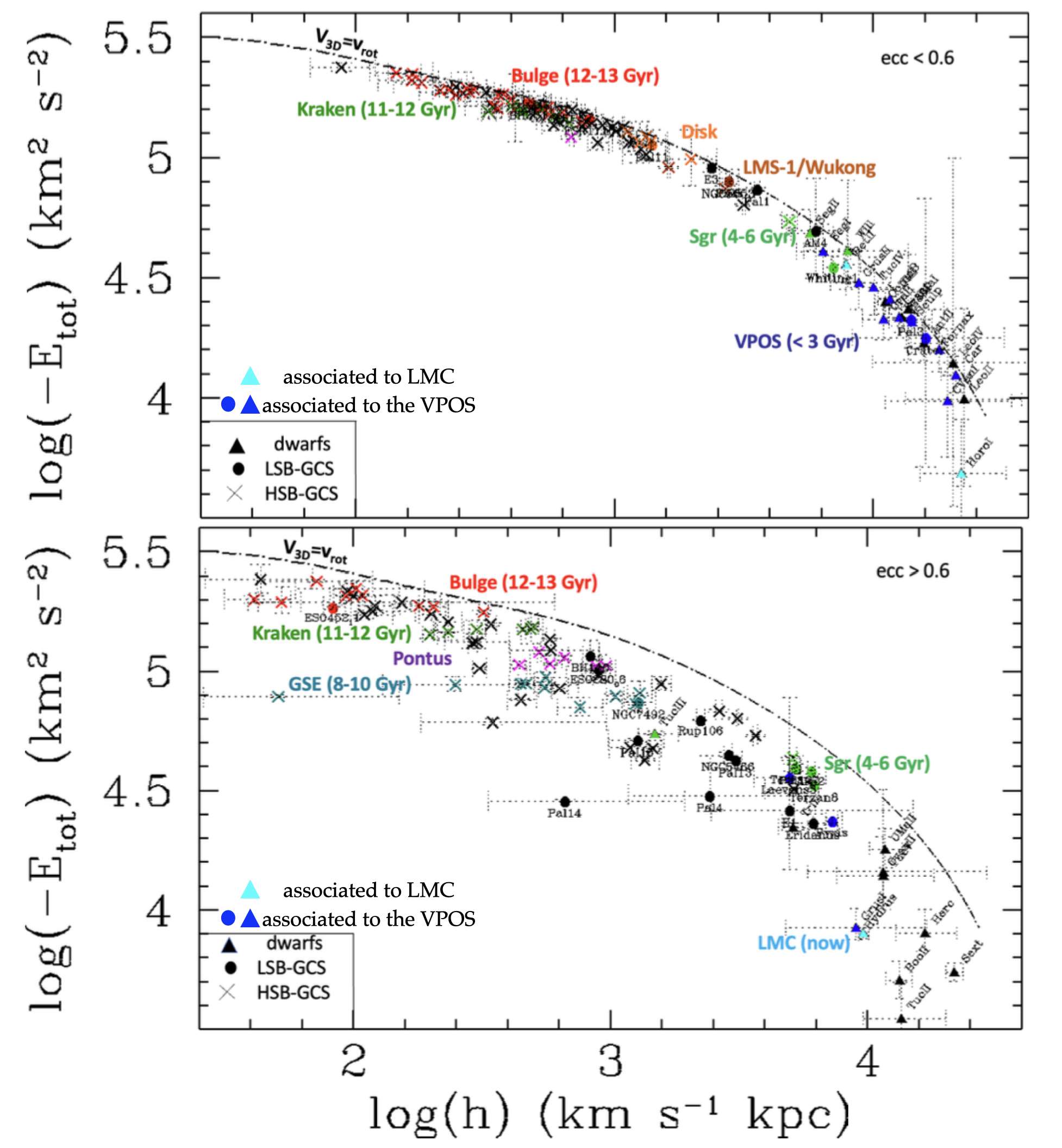}
  \caption{
  	The diagram of angular momentum ($h$) versus total energy for various samples, including high-surface brightness globular clusters (HSB-GCs) represented by crosses, low-surface brightness globular clusters (LSB-GCs) represented by filled circles, and dwarf galaxies represented by triangles. 
     The top and bottom panels show the objects with orbital eccentricities smaller than 0.6 (including 79 globular clusters and 20 dwarf galaxies) and larger than 0.6 (including 77 globular clusters and 15 dwarf galaxies), respectively. The dotted-dashed line represents the limit of the lowest energy (corresponding to a circular orbit) as a function of angular momentum.
  	The identified structures, as well as their first infall epochs, as determined by \citep{Malhan2022} and \citep{Kruijssen2020}, are indicated by different colors in the figure. VPOS dwarf galaxies and globular clusters are also shown in blue, while the few dwarf galaxies associated with the LMC are shown with cyan.
   	Note that few dwarf galaxies with positive energy are not shown, including Leo I in the VPOS , Carina II and III and Phoenix II, all related to the LMC, and Hydra II and Leo V.
}
  \label{fig:hetot}
\end{figure}

\begin{figure}[t]
  \centerline{\vbox to 6pc{\hbox to 10pc{}}}
  \centering
  \includegraphics[width=8cm]{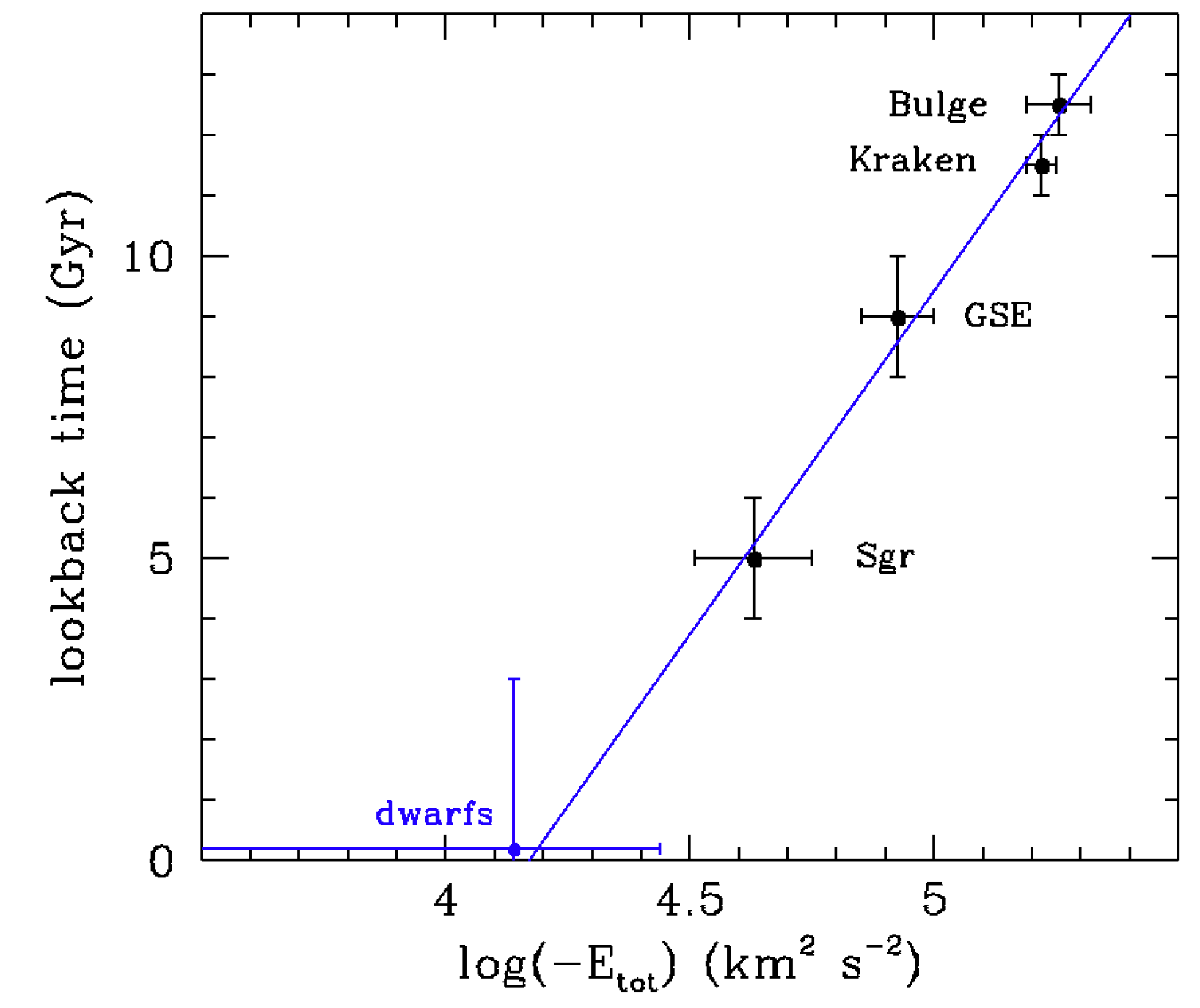}
  \caption{A relationship established from the analysis of Figure~\ref{fig:hetot} shows that the in-falling history of the MW's components can be accurately traced by their total orbital energy.}
  \label{fig:et}
\end{figure}

\subsection{High angular momenta and total energies of MW dwarf galaxies}
If MW dwarf galaxies are ancient inhabitants as it has been often assumed, we should be able to find such evidence by comparing them to the other MW inhabitants, such as globular clusters or halo stars. With fast growing observations, especially from Gaia, 
we compiled catalogues of complete 6D dynamical data for large samples of different tracers located at least 20 kpc from the MW center, such as K-gaint halo stars, globular clusters, and stars in the Sgr stream \citep[see][for more details]{Hammer2021}. Then, we compared the angular momenta and total energies of MW dwarf galaxies with those of the different tracers.
In Figure.~\ref{fig:hande} we present one of such comparisons, taking into account the completeness in volume among the different samples. It shows that the MW dwarf galaxies possess systematically high values in both angular momentum and total orbital energy, compared to other MW halo inhabitants. Both angular momentum and total energy are fundamental dynamical quantities, and are expected to be conserved over a long period of time if the system evolved  adiabatically. Even with possible dissipation of energy in the system, one may expect to see ordered evolutions in both angular momentum and total energy. Hence, the high angular momenta and total energies of MW dwarf galaxies indeed indicate that they have different dynamical properties than other MW halo populations.

\subsection{Relation between Infall time and total energy}
\label{sec:tandetot}
To further investigate the dynamical properties of MW dwarf galaxies, Figure.~\ref{fig:hetot} presents the relation between the total angular momentum ($h$) versus the total orbital energy ($E_{\rm tot}$), assuming a MW mass of $8.1\times10^{11}M_{\odot}$.  For comparison, we plot 156 globular clusters, stellar streams, and other known substructures of the MW \citep[see][for more details]{Hammer2023}. For a better visualisation, we separate the sample into those with eccentricities smaller (upper panel) and larger (lower panel) than 0.6.
Figure.~\ref{fig:hetot} suggests an impressively tight relation between angular momentum and total energy, particularly for cases with eccentricities below 0.6. 

\citet{Kruijssen2020} and \citet{Malhan2022} identified several substructures in the Galactic halo according to their dynamical properties, and  derived their accretion epochs, such as Kraken, Pontus, Gaia Sausage Enceladus (GSE), LMS1-Wukong, and Sgr. Using their results, we labeled each of these identified substructures associated with globular clusters in Figure.~\ref{fig:hetot}, along with their names and infall epochs. 
Interestingly, there appears to be a good correlation between the infall epoch and the total energy of substructures. The Large Magellanic Cloud (LMC) is widely accepted at the pericenter of its first infall to the MW \citep{Besla2007}, so it is labeled "now". From these results, we may infer that MW dwarf galaxies are likely latecomers, appearing later than the Sgr dSph whose first infall may have occurred about 4 Gyr ago \citep[e.g.,][]{Wang2022}. For a detailed analysis, we refer readers to \citet{Hammer2023}.

The relationship between infall epoch and total energy can be represented as in Figure.~\ref{fig:et}. The point labeled "dwarfs" is derived from 25 dwarf galaxies that do not belong to the low-energetic Sgr system (excluding also Sgr, Segue II, Tucana III and Willman I) or to the high energetic LMC system (Carina II, Carina III, Horologium I, Hydrus I, Phoenix II, and Reticulum II).  
The blue line in the figure is a linear fit to the data. A simple interpretation is that dwarf satellites with ${\rm log}(-E_{\rm tot}/[{\rm km}^2 {\rm s}^{-2}])  < 4.17$ are in their initial approach phase, with a value very close to the logarithm of the average energy (4.14) of 25 dwarf galaxies, whose scatter provides an upper limit of $-E_{\rm tot} = 4.34$. The latter combined with the linear fit suggests a lookback time of halo entry smaller than 3~Gyr. This possibly explains why MW dwarf galaxies show a strong tendency to be near to their pericenters. 

The relation in Figure.~\ref{fig:et} describes that the MW was likely to be formed by accretion and structured similarly to the onion skin model initiated by \citet{Gott1975} for explaining the outer density profile of elliptical galaxies. Together with the growth of the MW and possible energy dissipation mechanisms, such as dynamical frictions and tidal destruction of substructures, early infall components stay at lower energy orbits while later infall objects stay at higher energy orbits. The dynamical formation history in Figure.~\ref{fig:et} is consistent with the fact that the MW has not experienced major mergers since 10 Gyr \citep{Hammer2007,Haywood2018,Belokurov2018}, and also in excellent agreement with cosmological simulation results \citet{Rocha2012} and \citet{Boylan-Kolchin2013}.

\section{Possible consequences of the recent infall of MW dwarf galaxies}

Based on Gaia EDR3 data, we possess complete 6D dynamical data for MW dwarf galaxies. Our comprehensive analysis of their orbital properties suggests that most of them are likely to be the latecomers arrived in the MW within the last 3 Gyr. This finding completely changes the framework for understanding their origins because they were formerly assumed to be long-live inhabitants of the MW since 8--10 Gyr ago. 

Dwarf galaxies in the Local Group show the well known dichotomy that dwarf galaxies within about 250 kpc from the MW or M31 are typically devoid of gas, with only a few exceptional cases (which are relatively very massive objects) like the LMC, SMC, and IC10 \cite[e.g.,][]{Putman2021}. This phenomenon has been taken as one of the strong evidences that the diffuse gas in the Galactic halo has played a crucial role in the formation of dSphs, which are thought to have been transformed from dwarf irregulars (dIrrs) through ram-pressure stripping induced by the Galactic halo diffuse gas. This mechanism has been studied intensively and proved to be effective by \citet{Mayer2006}. However, according to our analysis of orbits, the time required to transform a dIrr into a dSph is within about 3 Gyr rather than the 10 Gyr as assumed by \citet{Mayer2006}. This suggests that a rapid and voilent ram-pressure stripping during the transformations from dIrrs to dSphs may have occurred, which may have had a substantial impact on their star formation, chemical evolution, and dynamical evolution.

One possible imprint of such violent transformations from dIrrs $\rightarrow$ dSphs may be breaks in the density profiles of dSphs that have been noticed since a long time \citep[e.g.,][]{Westfall2006}. 
At this epoch these breaks are detected only in classical dSphs. 
In Fornax, an extended stellar component was discovered recently by \citet{Yang2022}, which confirms the fact that all classical dSphs, except Leo~II, show a break in their density profiles. This indicates that a second and extended stellar component is common in the classical dSphs. Sgr is not taken into account here, as it has a different infall history compared to the other dSphs, see our analysis in Sect.~\ref{sec:tandetot}. 
In fact, an extremely large and faint stellar component has been identified recently using spectroscopy not only in the classcial dSph UMi \citep{Sestito2023}, but also in some of the UFDs: Ursa Major I, Coma Berenices, Bootes I \citep{Waller2023}, Tucana II \citep{Chiti2023}, and Grus I \cite{Cantu2021}.
These extended stellar components are found to be symmetrically surrounding the dSphs, indicative of their non-tidal origins. 
Could they possibly be the relics of the violent transformation from dIrrs $\rightarrow$ dSphs? For example, the removal of the last remaining gas may have caused a sudden lack of gravity, resulting in an expansion of the progenitor's core, and finally the formation of an extended stellar component. 
Such a possibility of destabilisation has been proposed by \citet{Yang2014}, see also \citep{Hammer2023}. 
In this context, the internal dynamics of the dSphs could also be affected, prompting us to reassess the estimates of their dynamical masses.

In brief, it is likely that most of the MW dwarf galaxies are latecomers that were accreted by the MW during the past 3 Gyr, as evidenced by their orbital properties revealed from the Gaia EDR3 data. This provides us with a new frame to understand how dIrrs have been transformed into dSphs in such a short period of 3 Gyr. For most of the MW dSphs, such a transformation just happened  during their first infall. 
We expect that future efforts will be made towards a precise determination of the star formation history, chemical evolution, structure and internal dynamics of dSphs, accounting in particular for a recent ram pressure stripping event associated to their revised orbital histories.


\begin{discussion}

\discuss{Ekta Patel}{
Have you computed the number of satellites expected to be at apocenter ? We have cosmological expectations for the total expected MW satellite population but these are certainly spread between pericenter and apocenter with a bias towards apocenter (at least from basic Keplerian arguments). One could do this for a range of MW masses.
}

\discuss{Yanbin Yang}{Thanks! We did such calculations and comparisons. As shown in Figure.~\ref{fig:rperi}, we tested the significance of the observed tendency to the results from a null hypothesis based on the Keplerian arguments, just as you have described for cosmological expectations (see text Sect.~\ref{sec:rperi}). }

\end{discussion}


\begin{thebibliography}{}
\bibitem[Belokurov et al.(2018)]{Belokurov2018} Belokurov, V., Erkal, D., Evans, N.~W., et al.\ 2018, \mnras, 478, 611
\bibitem[Besla et al.(2007)]{Besla2007} Besla, G., Kallivayalil, N., Hernquist, L., et al.\ 2007, \apj, 668, 949
\bibitem[Boylan-Kolchin et al.(2013)]{Boylan-Kolchin2013} Boylan-Kolchin, M., Bullock, J.~S., Sohn, S.~T., et al.\ 2013, \apj, 768, 140
\bibitem[Bullock \& Boylan-Kolchin(2017)]{Bullock2017} Bullock, J.~S. \& Boylan-Kolchin, M.\ 2017, \araa, 55, 343
\bibitem[Cantu et al.(2021)]{Cantu2021} Cantu, S.~A., Pace, A.~B., Marshall, J., et al.\ 2021, \apj, 916, 81
\bibitem[Cautun \& Frenk(2017)]{Cautun2017} Cautun, M. \& Frenk, C.~S.\ 2017, \mnras, 468, L41
\bibitem[Chiti et al.(2023)]{Chiti2023} Chiti, A., Frebel, A., Ji, A.~P., et al.\ 2023, \aj, 165, 55
\bibitem[Eilers et al.(2019)]{Eilers2019} Eilers, A.-C., Hogg, D.~W., Rix, H.-W., et al.\ 2019, \apj, 871, 120
\bibitem[Fritz et al.(2018)]{Fritz2018} Fritz, T.~K., Battaglia, G., Pawlowski, M.~S., et al.\ 2018, \aap, 619, A103
\bibitem[Gaia Collaboration et al.(2016)]{Gaia2016} Gaia Collaboration, Prusti, T., de Bruijne, J.~H.~J., et al.\ 2016, \aap, 595, A1
\bibitem[Gaia Collaboration et al.(2021)]{Gaia2021} Gaia Collaboration, Brown, A.~G.~A., Vallenari, A., et al.\ 2021, \aap, 649, A1
\bibitem[Gott(1975)]{Gott1975} Gott, J.~R.\ 1975, \apj, 201, 296
\bibitem[Ji et al.(2021)]{Ji2021} Ji, A.~P., Koposov, S.~E., Li, T.~S., et al.\ 2021, \apj, 921, 32
\bibitem[Jiao et al.(2021)]{Jiao2021} Jiao, Y., Hammer, F., Wang, J.~L., et al.\ 2021, \aap, 654, A25
\bibitem[Hammer et al.(2007)]{Hammer2007} Hammer, F., Puech, M., Chemin, L., et al.\ 2007, \apj, 662, 322
\bibitem[Hammer et al.(2021)]{Hammer2021} Hammer, F., Wang, J., Pawlowski, M.~S., et al.\ 2021, \apj, 922, 93
\bibitem[Hammer et al.(2023)]{Hammer2023} Hammer, F., Li, H., Mamon, G.~A., et al.\ 2023, \mnras, 519, 5059
\bibitem[\protect\citeauthoryear{Haywood et al.}{2018}]{Haywood2018} Haywood M., Di Matteo P., Lehnert M.~D., Snaith O., Khoperskov S., G{\'o}mez A., 2018, ApJ, 863, 113
\bibitem[Li et al.(2021)]{Li2021} Li, H., Hammer, F., Babusiaux, C., et al.\ 2021, \apj, 916, 8
\bibitem[Lindegren et al.(2021)]{Lindegren2021} Lindegren, L., Klioner, S.~A., Hern{\'a}ndez, J., et al.\ 2021, \aap, 649, A2
\bibitem[Lynden-Bell(1976)]{Lynden-Bell1976} Lynden-Bell, D.\ 1976, \mnras, 174, 695
\bibitem[Mayer et al.(2006)]{Mayer2006} Mayer, L., Mastropietro, C., Wadsley, J., et al.\ 2006, \mnras, 369, 1021
\bibitem[McConnachie(2012)]{McConnachie2012} McConnachie, A.~W.\ 2012, \aj, 144, 4
\bibitem[Malhan et al.(2022)]{Malhan2022} Malhan, K., Ibata, R.~A., Sharma, S., et al.\ 2022, \apj, 926, 107
\bibitem[Martin et al.(2008)]{Martin2008} Martin, N.~F., de Jong, J.~T.~A., \& Rix, H.-W.\ 2008, \apj, 684, 1075
\bibitem[Kruijssen et al.(2020)]{Kruijssen2020} Kruijssen, J.~M.~D., Pfeffer, J.~L., Chevance, M., et al.\ 2020, \mnras, 498, 2472
\bibitem[Kroupa et al.(2005)]{Kroupa2005} Kroupa, P., Theis, C., \& Boily, C.~M.\ 2005, \aap, 431, 517
\bibitem[Pawlowski et al.(2012)]{Pawlowski2012} Pawlowski, M.~S., Pflamm-Altenburg, J., \& Kroupa, P.\ 2012, \mnras, 423, 1109
\bibitem[Putman et al.(2021)]{Putman2021} Putman, M.~E., Zheng, Y., Price-Whelan, A.~M., et al.\ 2021, \apj, 913, 53
\bibitem[Riello et al.(2021)]{Riello2021} Riello, M., De Angeli, F., Evans, D.~W., et al.\ 2021, \aap, 649, A3
\bibitem[Riley et al.(2019)]{Riley2019} Riley, A.~H., Fattahi, A., Pace, A.~B., et al.\ 2019, \mnras, 486, 2679
\bibitem[Rocha et al.(2012)]{Rocha2012} Rocha, M., Peter, A.~H.~G., \& Bullock, J.\ 2012, \mnras, 425, 231
\bibitem[Sestito et al.(2023)]{Sestito2023} Sestito, F., Zaremba, D., Venn, K.~A., et al.\ 2023, arXiv:2301.13214
\bibitem[Simon(2019)]{Simon2019} Simon, J.~D.\ 2019, \araa, 57, 375
\bibitem[Torrealba et al.(2016)]{Torrealba2016} Torrealba, G., Koposov, S.~E., Belokurov, V., et al.\ 2016, \mnras, 459, 2370
\bibitem[Waller et al.(2023)]{Waller2023} Waller, F., Venn, K.~A., Sestito, F., et al.\ 2023, \mnras, 519, 1349
\bibitem[Walker(2013)]{Walker2013} Walker, M.\ 2013, Planets, Stars and Stellar Systems. Volume 5: Galactic Structure and Stellar Populations, 1039
\bibitem[Wang et al.(2022)]{Wang2022} Wang, H.-F., Hammer, F., Yang, Y.-B., et al.\ 2022, \apjl, 940, L3
\bibitem[Westfall et al.(2006)]{Westfall2006} Westfall, K.~B., Majewski, S.~R., Ostheimer, J.~C., et al.\ 2006, \aj, 131, 375
\bibitem[Yang et al.(2014)]{Yang2014} Yang, Y., Hammer, F., Fouquet, S., et al.\ 2014, \mnras, 442, 2419
\bibitem[Yang et al.(2022)]{Yang2022} Yang, Y., Hammer, F., Jiao, Y., et al.\ 2022, \mnras, 512, 4171
\end{thebibliography}
\end{document}